\documentclass[aip,pra,reprint,amsfonts,amsmath,amssymb,superscriptaddress,10pt,floatfix]{revtex4-1}

\usepackage{etex}
\reserveinserts{28}

\usepackage[english]{babel}
\usepackage[utf8]{inputenc}

\usepackage{amsmath}
\usepackage{amsfonts}
\usepackage{amssymb}
\usepackage{amsthm}
\usepackage{mathtools}
\usepackage{graphicx}
\usepackage{xfrac}
\usepackage{nicefrac}

\usepackage{color}
\usepackage{array}
\usepackage{longtable}
\usepackage{calc}
\usepackage{multirow}
\usepackage{hhline}
\usepackage{ifthen}

\usepackage[version=4]{mhchem}

\usepackage{times}
\usepackage{mathptmx}

\usepackage[colorlinks]{hyperref}

\graphicspath{{figures/}}

\hypersetup{citecolor=blue, urlcolor=blue, linkcolor=blue, colorlinks=true}

\begin{document}

\title{Solvent reaction coordinate for an S$_{\text{N}}$2 reaction}

\author{Christian Leitold}
\affiliation{Department of Chemical and Biomolecular Engineering, University of Illinois at Urbana-Champaign, 61801, USA}
\affiliation{Faculty of Physics, University of Vienna, 1090 Wien, Austria (present address)}

\author{Christopher J. Mundy}
\author{Marcel D. Baer}
\author{Gregory K. Schenter}
\affiliation{Physical Sciences Division, Pacific Northwest National Laboratory, Richland, WA 99352, USA}

\author{Baron Peters}
\email[Author to whom correspondence should be addressed. Electronic mail: ]{baronp@illinois.edu}
\affiliation{Department of Chemical and Biomolecular Engineering, University of Illinois at Urbana-Champaign, 61801, USA}
\affiliation{Department of Chemistry and Biochemistry, University of Illinois at Urbana-Champaign, 61801, USA}

\date{\today}

\begin{abstract}
We study the prototypical S$_{\text{N}}$2 reaction \ce{Cl- + CH3Cl -> CH3Cl + Cl-} in water using quantum mechanics / molecular mechanics (QM/MM) computer simulations with transition path sampling and inertial likelihood maximization. We have identified a new solvent coordinate to complement the original atom-exchange coordinate used in the classic analysis by Chandrasekhar, Smith, and Jorgensen.\cite{Chandrasekhar1985} The new solvent coordinate quantifies instantaneous solvent induced polarization relative to the equilibrium average charge density at each point along the reaction pathway. On the basis of likelihood scores and committor distributions, the new solvent coordinate improves upon the description of solvent dynamical effects relative to previously proposed solvent coordinates. However, it does not increase the transmission coefficient or the accuracy of a transition state theory rate calculation.
\end{abstract}

\maketitle

\section{Introduction}

The rates of many chemical reactions are dramatically altered by the presence of a polar solvent. The strongest solvent effects are typically observed for reactions that involve charge transfer and / or ion migration steps.\cite{Laidler1963, Reichardt2011} These reactions can be broadly classified into three categories: (i) electron transfer reactions where charges move between atoms with essentially fixed positions,\cite{Marcus1956, Small2003, VanVoorhis2010, Kuznetsov1999ElectronTI} (ii) ion association / dissociation reactions where atoms migrate while carrying their fixed charges,\cite{Onsager1938, Karim1986, Ciccotti1990, Dang1993,Geissler2001} and (iii) intermediate cases with concerted changes in the atom positions and atom charges.\cite{Reichardt2011, Hammes-Schiffer2010, Kretchmer2013, Warshel1980, Mayer2004, Knott2014} The discovery of vertical energy gap (solvent) reaction coordinates transformed our understanding of electron transfer processes,\cite{Marcus1956, Warshel1982} leading to many applications and generalizations.\cite{Albery1978, Marcus1993, nitzan2013chemical, Matyushov2007, Rasaiah2008} New interionic solvent density coordinates for ion association / dissociation\cite{Mullen2014a} have led to applications and generalizations in that domain.\cite{Yonetani2015, Roy2016, Joswiak2018, Joswiak2018a} In contrast to the widely applicable solvent coordinates for category (i) and (ii) reactions, solvent coordinates remain elusive for essentially all reactions in category (iii), including the S$_{\text{N}}$2, S$_{\text{N}}$1, E1, and E2 reactions.

The prototypical category (iii) reaction is the symmetric S$_{\text{N}}$2 “Finkelstein” reaction:
\begin{equation}
\ce{Cl- + CH_3Cl -> CH_3Cl + Cl-}.
\end{equation}
In a pioneering quantum mechanics / molecular mechanics (QM/MM) study of the Finkelstein reaction, Chandrasekhar, Smith, and Jorgensen\cite{Chandrasekhar1985} calculated the free energy profile along a difference of bond lengths,
\begin{equation}
r_J = r_{CCl(1)} - r_{CCl(2)}.
\end{equation}
where $r_{CCl(1)}$ and $r_{CCl(2)}$ are the two carbon--clorine bond distances. Free energy profiles computed with and without an aqueous solvent both show a barrier at $r_J = 0$, corresponding to the symmetric \ce{[Cl-CH3-Cl]-} transition state.\cite{Tirado-Rives2019} However, the aqueous solvent significantly increases the barrier, an equilibrium solvation effect that can be understood from basic electrostatics.\cite{HughesIngold} The solvent stabilizes the chloride ions in the reactant and product states more strongly than it stabilizes the delocalized charges in the transition state. Bash~et\,al.\cite{Bash1987} quantified these effects in combined semi-empirical + molecular mechanics simulations.  Specifically, they optimized the \ce{Cl- \bond{...} CH3Cl} configurations in the gas phase with different values of the \ce{Cl- \bond{...} C} distance. Then free energy perturbations at the fixed gas phase geometries with a molecular mechanics solvent showed how the solvent environment modified the activation energy and enhanced the equilibrium charges on the \ce{Cl- \bond{...} CH3Cl} atoms.

The free energy profile along a geometric solute coordinate like $r_J$ provides only limited information about the equilibrium solvation effects. To study the non-equilibrium dynamical effects, Bergsma et al.\cite{Bergsma1987} constructed an all-atom Hamiltonian from the QM/MM results of Chandrasekhar, Smith, and Jorgensen. Their Hamiltonian incorporated the potential of mean force along $r_J$ as well as Coulomb interactions between the solvent and solute atoms. In their Hamiltonian, the chlorine and carbon partial charges change as a prescribed function of $r_J$, with the charges at each $r_J$ taken from the solvent-averaged QM/MM results of Chandrasekhar, Smith, and Jorgensen. Thus their all-atom solute can polarize the solvent configurations, but the solvent configurations cannot induce polarization in the solute charge distribution. Gertner et al.\cite{Gertner1989} used the model show that the transmission coefficient for this system is near 0.5.


A small transmission coefficient for coordinate $r_J$ may result from inescapable friction in the dynamics,\cite{Kramers1940, Grote1980, Mullen2014a, Mullen2014} or from a sub-optimal definition of the reaction coordinate.\cite{vanderZwan1983, Pollak1986} As noted by earlier investigators,\cite{Pollak1986, Bergsma1987, Gertner1987, McRae2001, Ensing2004} an optimized reaction coordinate that includes solvent components may give a smaller friction, a higher transmission coefficient, and better predictions of the committor.  While several solvent coordinates have been investigated for the S$_{\text{N}}$2 system,\cite{Ensing2001} none have demonstrated an improvement over $r_J$ in terms of the transmission coefficient or committor analysis.

In this work, we use transition path sampling\cite{tps0} and inertial likelihood maximization\cite{Peters2012} to investigate one new solute coordinate and several new solvent coordinates.  Each new solvent coordinate accounts in some manner for solvent polarization.  One of the new solvent coordinates monitors instantaneous solvent-induced solute polarization, information that is not available from a prescribed charge-vs-$r_J$ relationship like that of Bergsma et al.\cite{Bergsma1987} Therefore, we use QM/MM trajectories with a classical SPC/Fw model\cite{Wu2006} for the water molecules, Gaussian electrostatic coupling,\cite{VandeVondele2005} and a Becke--Lee--Yang--Parr (BLYP) functional\cite{Becke1988,Lee1988a} for the methyl chloride and chloride ion.  After testing our simulation methods, we calculate the free energy as a function of the $r_J$ coordinate and compare $r_J$ to the new trial reaction coordinates.  The comparisons and coordinate optimizations are based on inertial log likelihood scores, transmission coefficients, and committor tests.

\section{Model and methods}
\label{sec:model_methods_finkelstein}

\begin{figure}
	\centering
	\includegraphics[width=0.75\columnwidth]{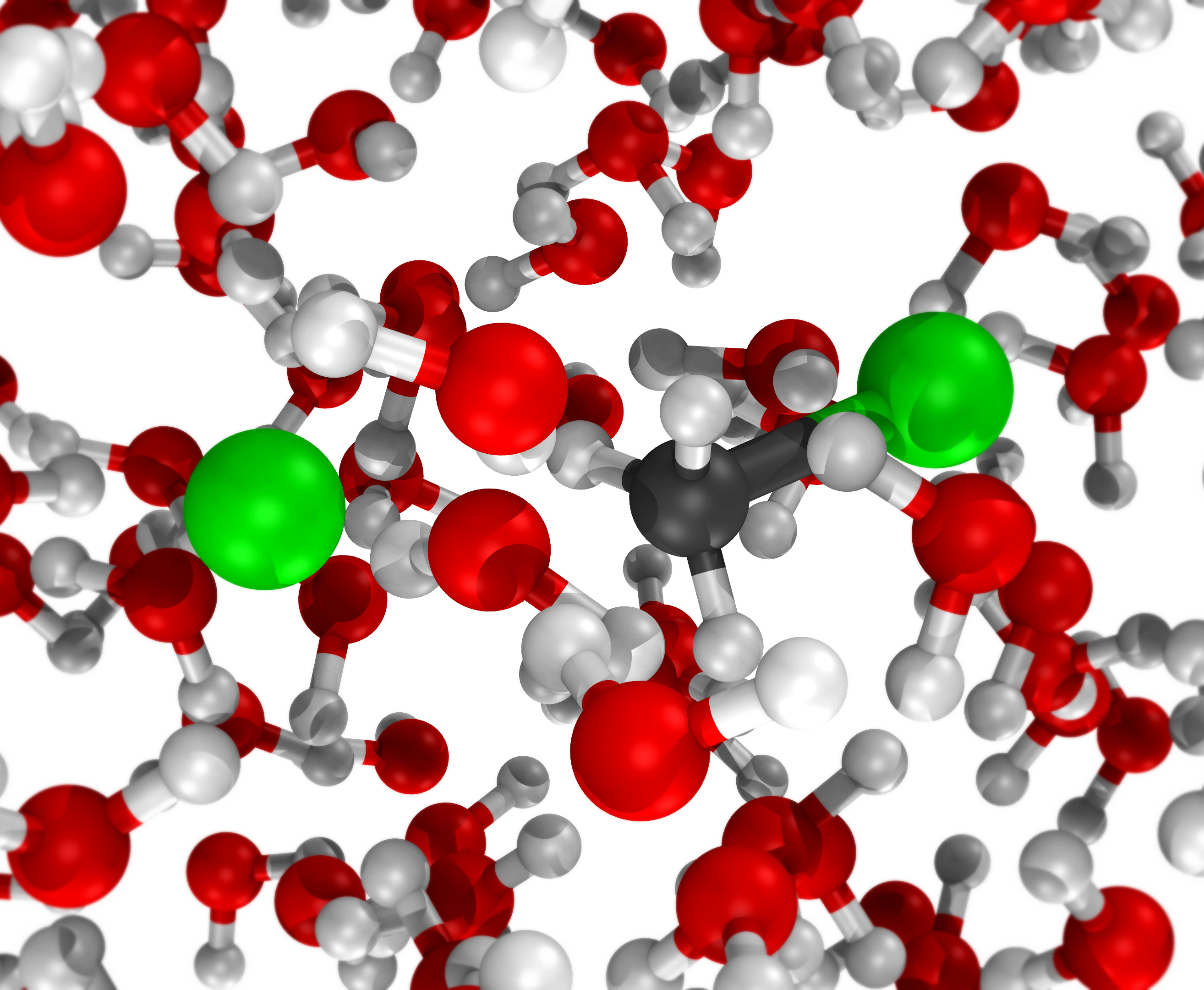}
	\caption{Rendering from our QM/MM simulation of the Finkelstein reaction in aqueous solution. The reaction is treated on the DFT level, while a classical model is employed for the solvent molecules.}
	\label{fig:rendering_modified}
\end{figure}

The methyl chloride molecule and the extra chloride ion are treated on the density functional theory (DFT) level using the Gaussian and plane waves (GPW) method.\cite{VandeVondele2005} We use a Gaussian polarized triple-zeta split-valence (TZV2P) basis set to expand the wave functions and the BLYP XC functional.\cite{Becke1988,Lee1988a} The electronic density is represented using an auxiliary plane-wave basis with an energy cutoff of 400\,Ry. Furthermore, we employ the pseudopotentials of Goedecker, Teter, and Hutter (GTH)\cite{Goedecker1996,Krack2005} to account for core electrons in C and Cl. The solvent, SPC/Fw water,\cite{Wu2006} is treated on the MM level. SPC/Fw is a simple flexible three-site water model with parameters tuned to reproduce the experimental values for the diffusivity and dielectric constant. Note that this water model, while having many positive attributes compared to
other common choices, lacks electronic polarization. While this is computationally very efficient, a more accurate description of the solvent might require a polarizable water model. OPLS-AA Lennard-Jones parameters\cite{Dodda2017} for methyl chloride are used to model the non-electrostatic part of the QM/MM cross interactions,
\begin{equation}
u_{ij}(r) = 4\varepsilon_{ij} \left[ \left(\frac{\sigma_{ij}}{r}\right)^{12} - \left(\frac{\sigma_{ij}}{r}\right)^{6} \right].
\end{equation}
Explicitly, the parameters are as follows, and we use Lorentz--Berthelot rules to obtain cross-species parameters.
\begin{center}
	\begin{tabular}{l|r|r}
	\hline
	\hline
	&$\varepsilon$ / kcal/mol&$\sigma$ / Å\\ \hline
	C &  0.066 & 3.5\\
	Cl & 0.300 & 3.4\\
	H & 0.030 & 2.4\\ \hline
	O & 0.1554253 & 3.165492\\
	\hline \hline
\end{tabular}
\end{center}
It is worth mentioning here that the LJ parameters for oxygen are the ones from the SPC/Fw water model, while there is no LJ center on the hydrogen atoms in SPC/Fw. These parameters are only employed for the non-electrostatic part of the QM/MM cross interactions, for example when an oxygen (present in the MM part only) interacts with a QM hydrogen. In practice, the interaction will be dominated by the electrostatic part, however, the repulsion from the Lennard-Jones potential is necessary to prevent unphysical overlaps between QM and MM atoms. A rendering of a simulation snapshot is shown in Fig.\,\ref{fig:rendering_modified}.

\subsection{Simulation details}
\label{subsec:simulation_finkelstein}

We perform QM/MM molecular dynamics simulations with an integration time step of $\Delta t = 0.5\,$fs implemented in CP2K. The package PYCP2K\cite{pycp2k} is used to run CP2K from a python interface. For both NVT and NpT simulations, we employ a Nosé--Hoover thermostat with a chain length of three. For the thermostat, the time constant is set to $\tau_T = 100\,$fs, while for the barostat we use $\tau_p = 1000\,$fs. In all our simulations, the system contains one methyl chloride molecule and a single chloride ion treated on the QM level, as well as an additional 505 SPC/Fw water molecules. Periodic boundary conditions are applied to the water molecules only, so we effectively simulate the reaction in the limit of infinite dilution. For the water molecules, smooth particle-mesh Ewald (SPME)\cite{Essmann1995} is used to compute the long-range part of the Coulomb interaction. The Ewald parameter is $\alpha = 0.44$ and 64 grid points are used in each direction of the Fourier grid. For all the short-ranged classical interactions, the potential is cut off at a radius of $r_{\text{cut}} = 9$\,Å.

Our TPS simulations are carried out at a temperature of 300\,K and a fixed volume of $V = 15.74$\,nm$^3$, which corresponds to the average volume of our umbrella sampling simulations performed at the same temperature and a pressure of 1\,bar. We check if one of the stable states is reached every 10 time steps, and we save the current snapshot along the path every 20 time steps. The shooting point is selected with equal probability from $\mathbf{x}_{0+\Delta T}^{(o)}$ and $\mathbf{x}_{0-\Delta T}^{(o)}$, where $\mathbf{x}_0^{(o)}$ denotes the shooting point of the previous trajectory and $\Delta T = 60 \, \Delta t$.\cite{Mullen2015}

\subsection{Umbrella sampling}

We perform a series of umbrella sampling simulations~\cite{TorrieValleauJCP1977} along Jorgensen's reaction coordinate $r_J$ in order to obtain the free energy. In the $n$-th of the equally spaced windows, the harmonic bias potential is given as
\begin{equation}
U_{\text{bias}}^n(r_J) = k \left(r_J - n \Delta r_J  \right)^2.
\end{equation}
We use a spring constant of $k = 20$\,kcal/mol/Å$^2$ and a window spacing of $\Delta r_J = 0.2$\,Å. The separate histograms were combined with WHAM\cite{GrossfieldWHAM} to obtain the probability density $p(r_J)$ and the free energy $F(r_J) = -k_B T \ln p(r_J)$. Since we also save the system configurations in regular intervals, we can later estimate the probability density as a function of alternative reaction coordinates $r_{\text{new}}$ by using conditional averages for a fixed value of $r_J$, i.\,e.
\begin{equation}
p(r_{\text{new}}) = \int dr_J \, p(r_{\text{new}} | r_J) \, p(r_J).
\end{equation}
Note that while exact in principle, in practice this will only yield reasonable results if $r_{\text{new}}$ is adequately and ergodically sampled at each value of $r_J$. Similarly, a two-dimensional probability density (and hence free energy) can be obtained by simply not performing the integration,
\begin{equation}
\label{eq:cond_prob}
p(r_J, r_{\text{new}}) = p(r_{\text{new}} | r_J) \, p(r_J).
\end{equation}

\subsection{Flexible path length transition path sampling with aimless shooting}

Transition path sampling (TPS) is a powerful technique for obtaining reactive dynamical pathways.~\cite{tps0,Bolhuis2002,tps1,Paul2019} We use TPS with aimless shooting~\cite{reactioncoords} and a flexible path length to sample trajectories crossing the reaction's barrier. Jorgensen's coordinate is used to distinguish between the two stable states, $A$ and $B$, and the transition region. A configuration is considered to be in state $A$ if $r_J \leq -2\,$Å and in state $B$ if $r_J \geq 2\,$Å. In the flexible path length version of TPS, no restrictions are placed on the length of the pathways. A path is simply accepted if it is reactive, i.\,e., $A \rightarrow B$ or $B \rightarrow A$, and rejected if it is not. As shown in Sec.\,\ref{sec:ilnmax}, we can later use information from both the accepted and the rejected pathways to optimize our reaction coordinate describing the transition.

\subsection{Inertial likelihood maximization}
\label{sec:ilnmax}

The aim of this study is to find a generic solvent reaction coordinate for chemical reactions in solution. In order to quantify if a proposed new coordinate is actually able to provide any improvement over existing coordinates, we use the method of inertial likelihood maximization.\cite{Peters2012} First, we start by defining a purely configurational trial reaction coordinate $q_T(\mathbf{x})$. For example, the trial reaction coordinate might be a simple linear combination of Jorgensen's solute coordinate $r_J$ and a new, yet to define solvent coordinate $q_p$:
\begin{equation}
\label{eq:linear_rc}
q_T(\mathbf{x}) = r_J(\mathbf{x}) + \alpha q_p(\mathbf{x}).
\end{equation}
In the next step, we introduce a committor-like model function, which is a function of the trial reaction coordinate and the velocity along the coordinate:
\begin{equation}
\tilde{p}_B(\mathbf{x}, \dot{\mathbf{x}}) \approx \frac{1}{2} \left\{1 + \text{erf}[a q_T(\mathbf{x}) + b \dot{q}_T(\mathbf{x}, \dot{\mathbf{x}})]\right\}.
\end{equation}
Note that by setting $b=0$ we can recover the original, inertia-free formulation of the likelihood maximization method.\cite{peters_rc} The actual committor $p_B(\mathbf{x})$ does only depend on the configuration, but not its velocity. Velocity contributions can be included in the definition of the dividing surface,\cite{Garcia-Meseguer2019} but such definitions are inconvenient for constructing simple rate expressions. The next step is to write down the expression for the likelihood of the observed simulation results based on the hypothesized trial reaction coordinate:
\begin{equation}
\ln L = \sum_{\mathbf{x}^{(k)}}^{\rightarrow A} \ln \left[1 - \tilde{p}_B(\mathbf{x}^{(k)}) \right] + \sum_{\mathbf{x}^{(k)}}^{\rightarrow B} \ln \tilde{p}_B(\mathbf{x}^{(k)}).
\end{equation}
Here the set of points $\mathbf{x}^{(k)}$ are the shooting points from a TPS simulation using aimless shooting. The first sum runs over all trajectories ending in state $A$, while the second sum runs over all trajectories ending in $B$. Note that by including the velocity information along the model reaction coordinate, we can make use of both the forward and the backward segment of our shooting trajectories, effectively doubling the number of data points we get from each trajectory. Care must be taken to use the reversed velocities in the case of the backward trajectories. Finally, the (logarithmic) likelihood is optimized, in other words, one finds the set of parameters $(a, b, \alpha_1, \alpha_2, \dots, \alpha_M)$ such that $\ln L$ is maximal.

We performed additional committor tests for a number of trial coordinates that had high likelihood scores.\cite{tps0} Estimates for the true distribution of committor values have been obtained by deconvoluting the discrete committor histograms.\cite{Peters2006a}

\subsection{Transmission coefficients}

Transition state theory (TST) assumes that a trajectory crossing the dividing surface always continues on to the product state, i.\,e. there are no recrossings.\cite{Wigner1938, Chandler1978, BaronBookChapter13} For real systems trajectories can recross the dividing surface because of friction with the solvent or because the dividing surface has not been ideally positioned. These recrossing events always cause the true rate to be smaller than the TST estimate. The transmission coefficient is the ratio of the true rate and the TST estimate. For a reaction with a single barrier, an accurate reaction coordinate should maximize the transmission coefficient.

An efficient way to calculate the transmission coefficient is the method of effective positive flux or EPF.\cite{VanErp2012} In EPF, one first obtains a sample of initial points $(\mathbf{x}_0, \dot{\mathbf{x}}_0)$ from the top of the free energy barrier. Then, a point is considered \textit{positive} (P) if it has a positive velocity along the reaction coordinate, $\dot{r}(\mathbf{x}_0, \dot{\mathbf{x}}_0) > 0$. The point is considered \textit{first} (F) if the \textit{backwards} trajectory fragment, launched from $(\mathbf{x}_0, -\dot{\mathbf{x}}_0)$, reaches state $A$ without ever crossing back to the $B$-side of the barrier. Finally, a point is considered \textit{effective} (E) if the \textit{forwards} trajectory fragment reaches state $B$, and the backwards trajectory fragment reaches $A$. The transmission coefficient can be calculated as
\begin{equation}
\kappa(r) = \frac{\langle \dot{r}_0 \chi_{AB}^{EPF}(\mathbf{x}_0, \dot{\mathbf{x}}_0) \rangle_{\ddagger}}{\frac{1}{2} \langle |\dot{r}| \rangle_{\ddagger}},
\end{equation}
where $\chi_{AB}^{EPF}$ is the indicator function for the three conditions and $\langle \dots \rangle_{\ddagger}$ denotes an average over configurations on the top of the barrier.

As noted by van Erp,\cite{VanErp2012} one tests the conditions in the order P, F, and E to maximize computational efficiency. We select configurations within a very small interval around the barrier value of the respective coordinate from our umbrella sampling simulation. As these simulations are performed at constant $NpT$, the simulation box size will slightly vary from configuration to configuration. However, for each trajectory we keep the box size fixed and integrate the system at constant $NVT$.

\subsection{Committor analysis}
\label{sec:committor}

For any configuration $\mathbf{x}$, the committor $p_B(\mathbf{x})$ is the fraction of dynamical pathways started with random momenta from $\mathbf{x}$ that first reaches state $B$ before reaching state $A$.\cite{tps0} In a committor test of a reaction coordinate $r$, one computes the distribution of estimated committor values for configurations launched from the top of the barrier according to the coordinate in question, $r = r^{\ddagger}$. A good reaction coordinate should be sufficient to predict the committor, i.\,e. all configurations on an isosurface of the coordinate should have similar committor values. A standard test for reaction coordinate accuracy generates a histogram of $p_B$ estimates for configurations on the trial coordinate isosurface  $r = r^{\ddagger}$. Mathematically the histogram is a convolution of the binomial distribution and the true committor distribution $\rho(p_B | r = r^{\ddagger})$. The mean and standard deviation $(\mu, \sigma)$ of this distribution and the histogram are related by\cite{Peters2006a}
\begin{align}
\mu &= \mu_H,\label{eq:mu}\\
\sigma &= \sqrt{\sigma_H^2 - \mu_H(1-\mu_H)/N},\label{eq:sigma} 
\end{align}
where $N$ is the number of trajectories used for each committor estimation and $(\mu_H, \sigma_H)$ are the mean and standard deviation of the histogram of estimated committors. For the S\textsubscript{N}2 reaction with its single barrier, a good reaction coordinate and ideally placed dividing surface should give a mean of $1/2$ and a standard deviation as small as possible. This provides us with a method for comparing the quality of different proposed reaction coordinates.

\section{Trial coordinates}
\label{sec:trialcoordinates}

\subsection{Polarization-based solvent reaction coordinate}

The primary contribution of this work is the introduction of a new solvent reaction coordinate, which is applicable to a wide range of chemical reactions in solution. Specifically, we monitor the solvent polarization by its influence on the instantaneous charges on the solute atoms. For each sampled configuration, we use Mulliken population analysis\cite{Mulliken1955} to assign charges to each QM atom, but our proposed solvent coordinate is generic enough that the exact details of the charge assignment should not matter. For each fixed value of the solute coordinate $r_s$, these charges will have a well-defined average. Conversely, a random instantaneous configuration at the same value of $r_s$ will in general have charges that deviate from the corresponding average. It is exactly this fluctuation from the average which forms the basis of the new coordinate.

\begin{figure}
	\centering
	\includegraphics[width=0.90\columnwidth]{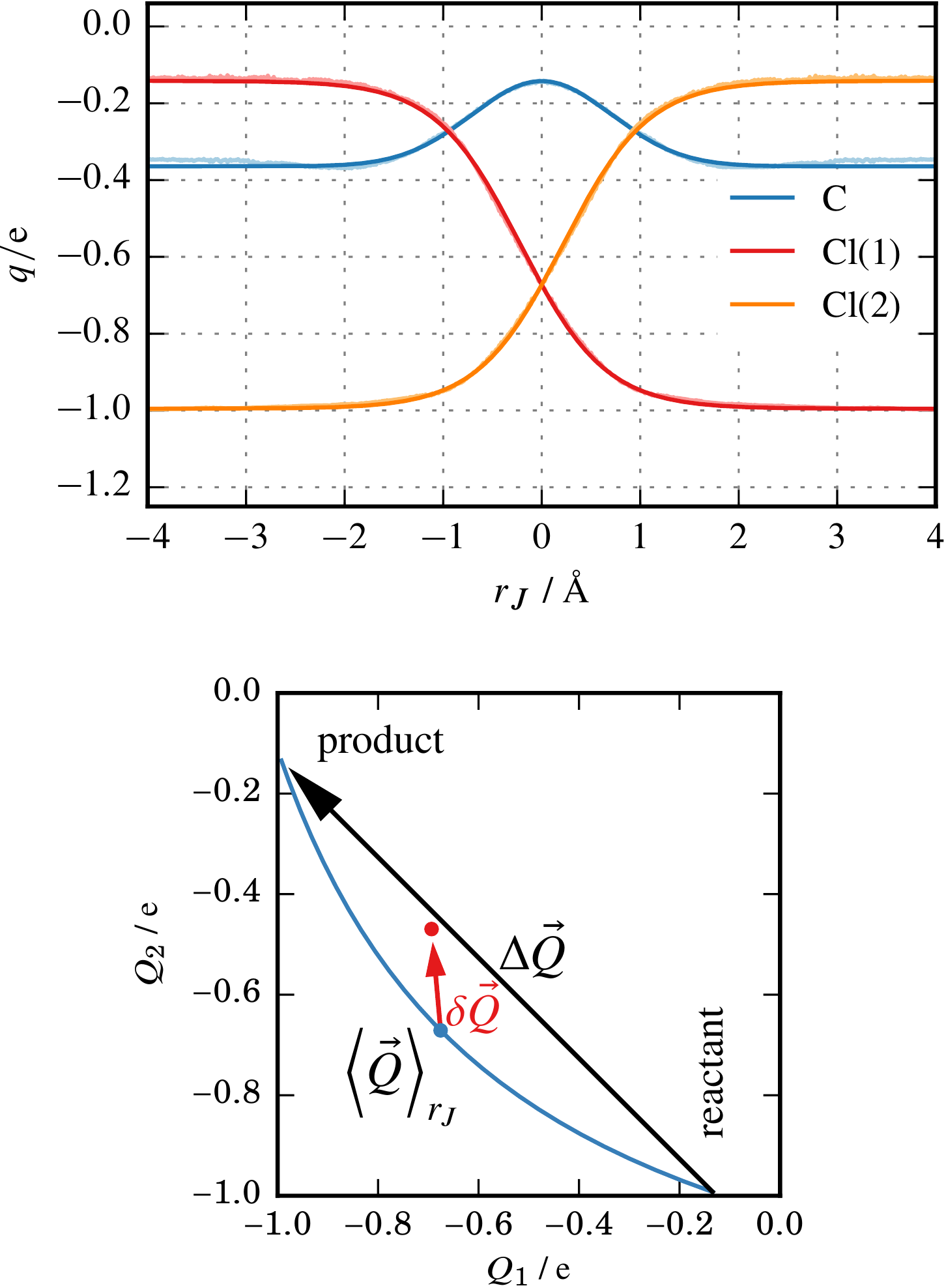}
	\caption{(Top) Average Mulliken charge on the chlorine atoms and the central carbon atom as a function of Jorgensen's coordinate. The darker solid lines are fit functions to the actual data plotted below. Note that with our convention for the definition of $r_J$, the incoming atom is denoted as Cl(2). (Bottom) Construction of the new solvent reaction coordinate $q_p$. The blue line shows the average progress of charges on the two chlorine atoms as the solute coordinate marches from the reactant to the product state. Individual trajectories will not necessarily follow the blue path exactly. The red circle shows the charges for an instantaneous solvent configuration, which in general will deviate from the averages indicated by the blue circle.}
	\label{fig:solvent_rc_construction}
\end{figure}

Let $\vec{Q} = (Q_1, Q_2, \dots, Q_M)$ be the vector of charges on the $M$ solvent atoms. Then, in the space of charges,
\begin{equation}
\Delta \vec{Q} = \left \langle \vec{Q}\right \rangle_{\text{product}}  - \left \langle \vec{Q} \right \rangle_{\text{reactant}}
\end{equation}
is the vector pointing along the general direction of the reaction. Here, $\langle \vec{Q} \rangle_{\text{product}}$ (or $\langle \vec{Q} \rangle_{\text{reactant}}$) is the equilibrium average charge at the fixed value of the solute coordinate which corresponds to the product (or reactant) state. Similarly, for any configuration with a solute reaction coordinate of $r_s$ we can define
\begin{equation}
\delta \vec{Q} = \vec{Q} - \left \langle \vec{Q}\right \rangle_{r_s},
\end{equation}
where $\langle \vec{Q} \rangle_{r_s}$ is the equilibrium average charge at a fixed solute coordinate value of $r_s$. Hence, $\delta \vec{Q}$ is the instantaneous deviation from expected solute charge distribution due to instantaneous solvent configuration. Now, the question is whether this deviation points forward or backward along the reaction pathway. We can use a scalar product to quantify this question:
\begin{equation}
\label{eq:solvent-rc}
q_p = \delta \vec{Q} \cdot \Delta \vec{Q}.
\end{equation}
When the dot product is positive, the solvent has a product-like polarization, and when the dot product is negative, the solvent has a reactant-like polarization. The equation provides a very generic way of constructing a polarization-based solvent coordinate for reactions in solution. No particular details of the reaction in question are required, as long as we have an accurate solute coordinate and a way of assigning charges to atoms. Also note that the general strategy for construction of the solvent coordinate in Eq.\,\eqref{eq:solvent-rc}, naturally conforms to the free energy models in Eq.\,2.1 of work by Gertner et al.\cite{Gertner1987} Specifically, to compute the solute coordinate we subtract as an instantaneous solvent characteristic (in this case a solvent-affected solute characteristic) from its equilibrium constrained average at the frozen solute coordinate position. Accordingly, our free energy surfaces are expected to resemble those sketched in the theoretical analysis of Gertner et al. We show the construction of the solvent reaction coordinate in Fig.\,\ref{fig:solvent_rc_construction}.

\subsection{Other trial coordinates}

\begin{figure}
	\centering
	\includegraphics[width=0.8\columnwidth]{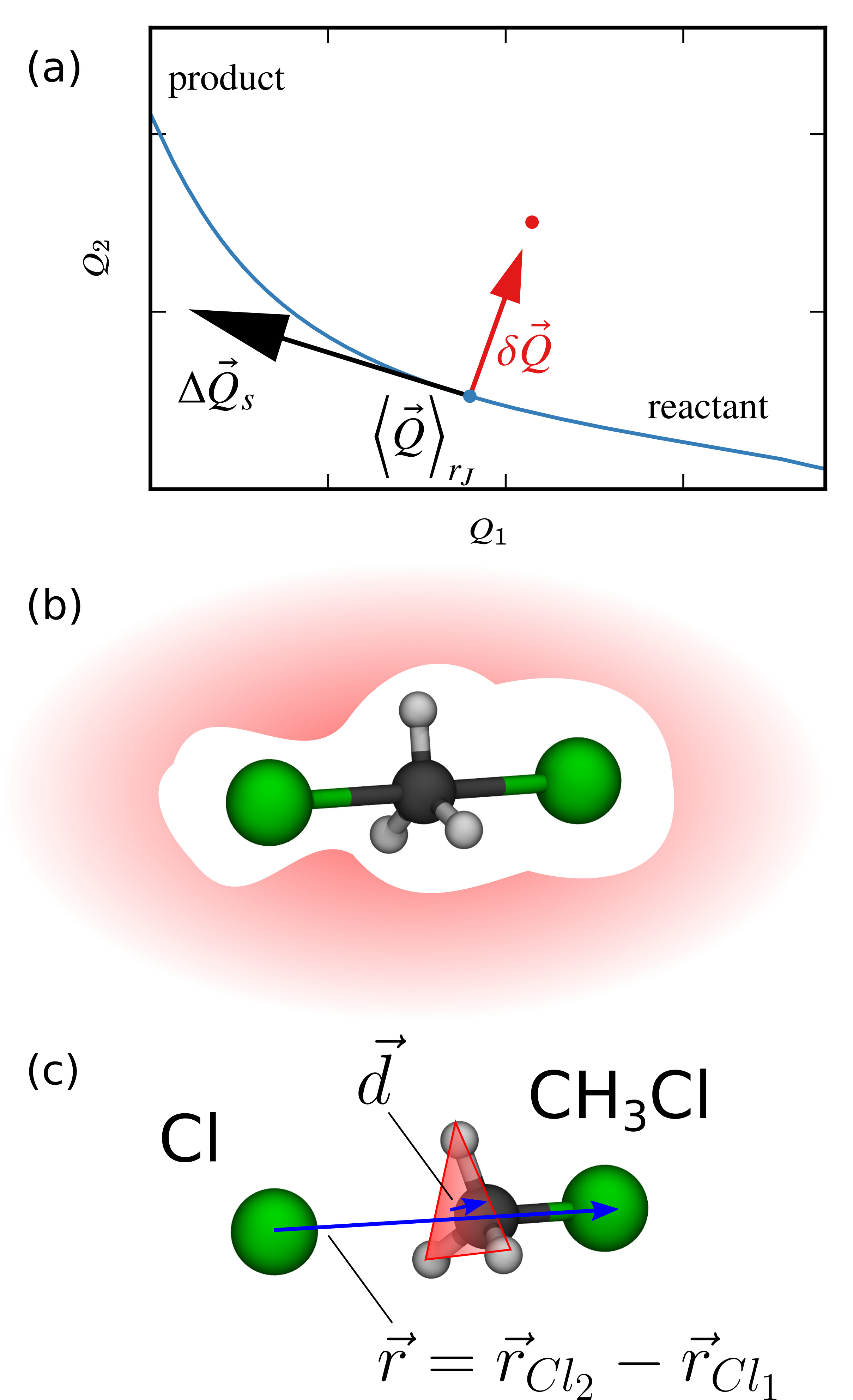}
	\caption{(a) Alternative polarization coordinate. (b) Pinching coordinate. The shaded red surrounding the molecule represents the solvent, which is “pinched in” more on one side of the central carbon than on the other. (c) Plane-inversion solute coordinate. The plane depicted in red is defined by the three hydrogen atoms. The vector $\vec{d}$ is perpendicular to the plane and points from the plane to the carbon atom. The plane-inversion coordinate $r_p$ is the scalar product of this vector with the vector pointing from the first to the second chlorine atom, normalized by the length of the latter.}
	\label{fig:finkelstein_three_coords}
\end{figure}

\subsubsection{Variant of the polarization solvent coordinate}

In a slight variation of the polarization coordinate, it is possible to use an alternative definition for the vector pointing in the direction of the reaction. While $\Delta \vec{Q}$ is the straight-line connection from reactant to product state in the charge space, one can instead use the smoothly varying derivative of the average charge as a function of the primary coordinate which is held fixed:
\begin{equation}
\Delta \vec{Q}_s(r_s) = R_s \frac{d}{d r_s} \left \langle \vec{Q} \right \rangle_{r_s},
\end{equation}
where $R_s$ is a normalization constant to ensure that $\Delta \vec{Q}_s$ has the same dimension and magnitude as the original $\Delta \vec{Q}$. Conversely, the alternative version of the solvent coordinate is calculated as $q_{ps} = \delta \vec{Q} \cdot \Delta \vec{Q}_s$. In this case, the results are very similar to the results for $q_p$. The construction is illustrated in Fig.\,\ref{fig:finkelstein_three_coords}(a), further details are provided in the Supplementary Material (SM).

\subsubsection{Pinching coordinate}

While this coordinate is also probing the effect of the solvent on the reacting molecule, it is purely based on coordination numbers and hence does not require any knowledge of atomic charges. The idea is to quantify how much the solvent cloud is “pinched in” on both sides of the central carbon, illustrated in a cartoon in Fig.\,\ref{fig:finkelstein_three_coords}(b). We start by defining two virtual sites $\vec{v}_1$ and $\vec{v}_2$, which are located at the midpoints between the central carbon and the first and second chlorine, respectively. Then, a coordination number is defined for each virtual site:
\begin{equation}
\label{eq:coordination_number}
c_{i,T} = \sum_j \frac{1 - \left(\frac{r_{ij}}{r_0}\right)^n}{1 - \left(\frac{r_{ij}}{r_0}\right)^{2n}},
\end{equation}
where the sum goes over all solvent particles $j$ of type $T$ in the vicinity of virtual site $i$ and $r_{ij}$ is the distance to the virtual site. The parameters $r_0$ and $n$ determine the range and steepness of the cutoff function. In practice, we only use the coordination number with respect to hydrogen atoms that are part of solvent water molecules. Then, we can define the pinching coordinate
\begin{equation}
\label{eq:pinching_coordinate}
\Delta c_H = c_{1,H} - c_{2,H}
\end{equation}
as the difference in the coordination numbers of the two virtual sites. Values for the parameters $r_0$ and $n$ can be determined using some suitable optimization procedure.

\subsubsection{Plane-inversion solute reaction coordinate}

Jorgensen's original reaction coordinate $r_J$ is the difference between the distances of two chlorine atoms from the central carbon atom. It  clearly separates the reactant and the product state. However, $r_J$ still leaves some room for improvement in the immediate vicinity of the transition state. In particular, $r_J$ is not sensitive to a Walden inversion\cite{Rordam1928} of the hydrogen atoms, which happen on a shorter timescale than the typical changes in $r_J$. In order to address this issue while still maintaining the symmetry property, we introduce a new solute coordinate. The coordinate is based on the directed distance of the central carbon atom from the plane defined by the three hydrogen atoms. We define the plane-inversion coordinate as
\begin{equation}
r_p = \frac{\vec{d} \cdot \vec{r}}{|\vec{r}|},
\end{equation}
where $\vec{r} = \vec{r}_{Cl_2} - \vec{r}_{Cl_1}$ and $\vec{d}$ is the vector normal to the plane and pointing from the plane to the carbon atom. The geometry is illustrated in Fig.\,\ref{fig:finkelstein_three_coords}(c), and the full functional expression for $\vec{d}$ is given in Appendix\,\ref{ap:plane-rc}.

\section{Results and discussion}
\label{sec:results_finkelstein}

\subsection{Equilibrium solvation structure}
\label{sec:finkelstein_solvation}

\begin{figure}
	\centering
	\includegraphics[width=1.0\columnwidth]{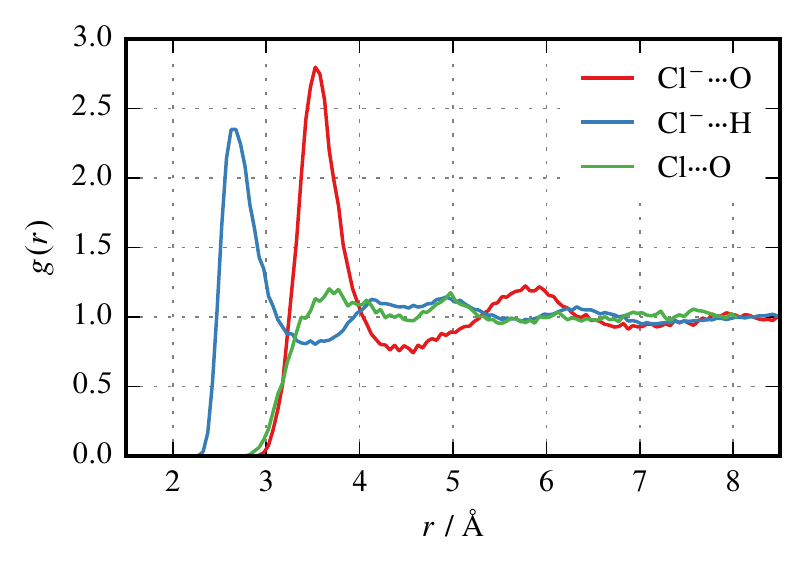}
	\caption{QM/MM pair correlation functions for the single chloride ion, as well as for the methyl chloride Cl, in SPC/Fw water.}
	\label{fig:gofr_chloride}
\end{figure}

As a first test of the validity of our QM/MM simulation, we have investigated the equilibrium solvation structure of a single QM chloride ion as well as a single QM methyl chloride molecule in SPC/Fw water. The simulations were performed at a temperature of 300\,K and a pressure of 1\,bar. The results for the pair correlation function $g(r)$ are shown in Fig.\,\ref{fig:gofr_chloride}. The location of the first peak in the chloride--oxygen pair correlation is in good agreement with results from Sala, Guàrdia, and Masia.\cite{Sala2010}

\subsection{Charge distribution and free energy along $r_J$}
\label{sec:finkelstein_charges}

In Fig.\,\ref{fig:solvent_rc_construction}, we show how the average charges change along the progress of the reaction. At the transition state, there is less negative charge on the central carbon atom. The charges on the hydrogen atoms, which are approximately constant along the reaction ($q_H / \text{e} \approx 0.16$), are not shown.

\begin{figure}
	\centering
	\includegraphics[width=1.0\columnwidth]{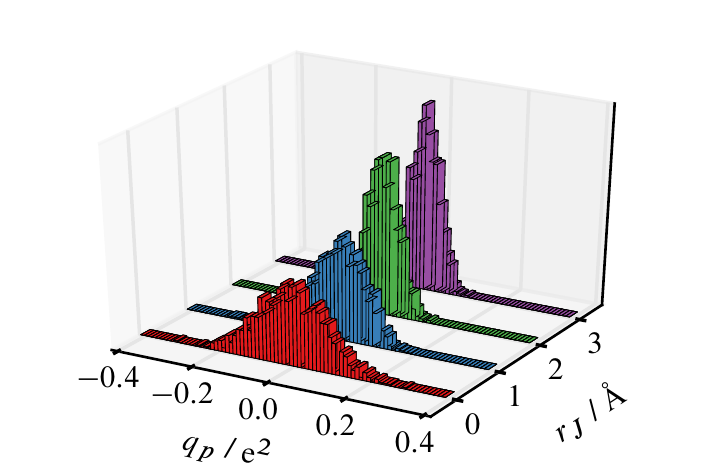}
	\caption{Distribution of the solvent coordinate $q_p$ for selected values of $r_J$. The distribution gets narrower if one moves away from the transition state.}
	\label{fig:qhist3d_new}
\end{figure}

The polarizability of the \ce{[Cl\bond{...}CH3\bond{...}Cl]^-} complex also increases at the transitions state. In Fig.\,\ref{fig:qhist3d_new}, we show the distribution of $q_p$ of different values of $r_J$. At the transition state, the fluctuations in the charges, and hence in $q_p$, are rather large. By construction, the distribution of $q_p$ at each $r_J$ is centered around $q_p = 0$.

\begin{figure}
	\centering
	\includegraphics[width=1.0\columnwidth]{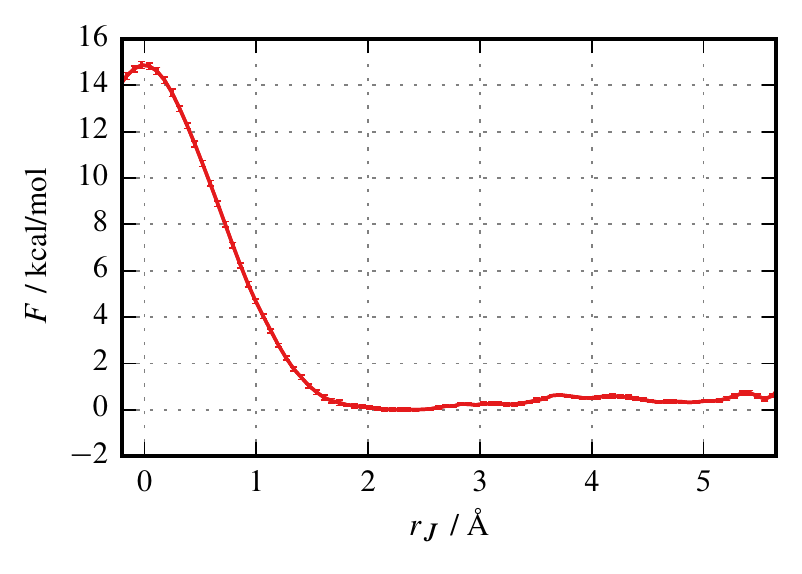}
	\caption{Free energy at $T = 300$\,K and $p = 1$\,bar from umbrella sampling along $r_J$. Error bars are calculated using Monte Carlo bootstrap error analysis implemented in WHAM.}
	\label{fig:new_fe}
\end{figure}

The one-dimensional free energy along Jorgensen's coordinate $r_J$ is shown in Fig.\,\ref{fig:new_fe}. The calculated barrier height of 15\,kcal/mol is much smaller than the estimated experimental barrier of 26\,kcal/mol. The small error bars suggest that the discrepancy is not due to the umbrella sampling procedure. Instead, from preliminary gas-phase calculations, we have reason to believe that the choice of exchange-correlation functional has a strong impact on the height of the barrier. Further details on that are provided in the SM. Although the absolute barrier height depends strongly on the chosen density functional, the barrier crossing dynamics (the focus of this work) may not. On the other hand, the imaginary frequency at the barrier top will change as the barrier height changes, and the imaginary frequency is known to influence transmission coefficients from Grote--Hynes theory\cite{Grote1980} and from prior simulation work.\cite{Gertner1989} Unfortunately, we were unable to examine transmission coefficients for more accurate hybrid functionals because of the QM/MM simulation cost. We have only performed calculations with BLYP because the same analysis with hybrid functionals or higher theory levels like CCSD(T) would have required far more CPU cores than we had available.

\subsection{Reaction coordinates}
\label{sec:finkelstein_rc}

We seek a better understanding of the reaction dynamics, particularly the non-equilibrium solvation dynamics. The trial reaction coordinates examined in this section include solvent coordinates like coordination numbers that have been used in previous studies, solvent coordinates that we developed (often inspired by earlier discussions), and alternative solute coordinates like a Walden inversion coordinate. This section presents likelihood rankings for various trial coordinates. For those which prove interesting in some way, we further investigate two-dimensional free energy surfaces, committor distributions, and transmission coefficients. 

In Table~\ref{tab:likelihoods} we show the optimized inertial log likelihood scores for a selected combination of reaction coordinates. Two- and three-variable coordinates are linear combinations analogous to Eq.\,\eqref{eq:linear_rc}. For comparing different coordinates, it is useful to employ a Bayesian criterion.\cite{bic} In particular, when increasing the model complexity by adding an additional parameter, the increase in the likelihood, $\Delta \! \ln L = \ln L - \ln L_{\text{ref}}$, should be at least
\begin{equation}
\Delta_{\text{min}} = \frac{1}{2} \ln n,
\end{equation}
where $n$ is the number of observations, which is twice the number of trajectories in the case of inertial likelihood maximization.

\begin{table}[htbp]
	
\centering
\setlength{\tabcolsep}{9pt} 
\def\arraystretch{1.3}	
\begin{tabular}{l|r r r}
\hline
\hline
\rule{0pt}{1.2\normalbaselineskip}
& \multicolumn{1}{l}{$\ln L$} & \multicolumn{1}{l}{$\Delta \! \ln L / \Delta_{\text{min}}$} & \multicolumn{1}{r}{$\kappa$}\\[1.5ex]
\hline
$r_p$, $r_J$, and $q_p$ & -1159.5 & 23.56 & -\\
$r_J$ and $q_p$ & -1178.2 & 19.04 & $0.32 \pm 0.07$\\
$r_p$ and $r_J$ & -1237.0 & 4.92 & $0.33 \pm 0.08$\\
$r_J$ only & -1257.5 & 0.00 & $0.39 \pm 0.07$\\
$r_p$ only & -1269.2 & -2.82 & $0.32 \pm 0.05$\\
\hline
$q_p$ only & -2743.2 & -357.17 & $0$\\
\hline \hline
\end{tabular}

\caption{Optimized inertial log likelihood scores for the different simple candidate reaction coordinates, and the corresponding transmission coefficient $\kappa$ of the coordinates for which we have calculated it. $\Delta\!\ln L$ is calculated with respect to using Jorgensen's coordinate $r_J$ only.\\
}
\label{tab:likelihoods}
	
\end{table}

Likelihood scores for additional variable combinations, including one containing the pinching coordinate $\Delta c_H$, are shown in the SM. Note that we have included two variants of the solvent coordinate as defined in Eq.\,\eqref{eq:solvent-rc}. Specifically, $q_p$ is parametrized with respect to fixing the original solute coordinate $r_J$, while $q_p'$ uses the new solute coordinate $r_p$ as reference. In the latter case, we have restricted the charge averaging as a function of $r_p$ to configurations in the vicinity of the transition state. Similarly, for $q_{ps}$ the charge parametrization is done with respect to $r_J$, while it is with respect to $r_p$ for $q_{ps}'$. For some of the coordinates, we have calculated the transmission coefficient $\kappa$ as well. Given that the transmission coefficient for $r_J$ is already quite high, we can at the most expect a slight improvement for alternative coordinates. Errors in $\kappa$ are estimated by simply performing a number of calculations for independent subsets of all transition states, and then averaging the results.

Two important messages are to be taken from the likelihood maximization results. First, presumably due to its high sensitivity right at the transition state, the plane-inversion coordinate performs comparable to Jorgensen's coordinate in predicting the outcome of dynamical trajectories. Second, the addition of the solvent coordinate $q_p$ or its variants to the model is able to further improve the log likelihood by about 20 times the value of $\Delta_{\text{min}}$. 

\begin{figure}
	\centering
	\includegraphics[width=1.0\columnwidth]{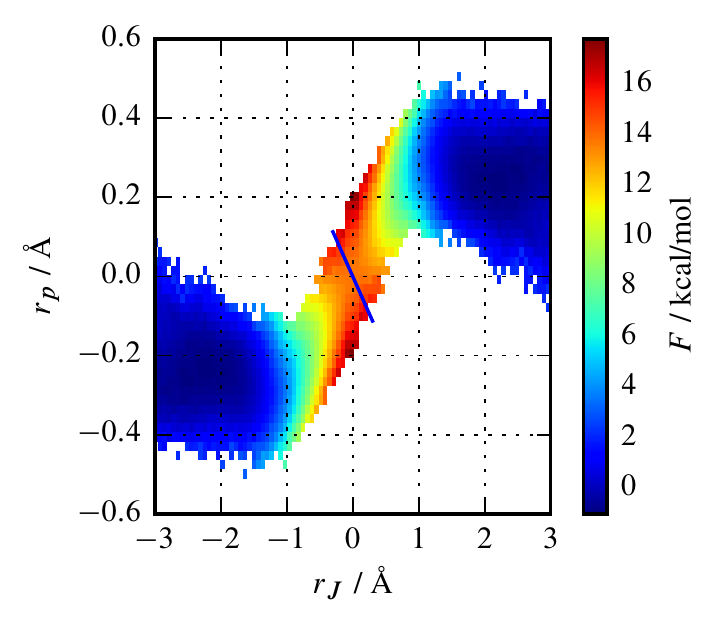}
	\caption{Free energy as a function of Jorgensen's coordinate $r_J$ and the new plane-inversion coordinate $r_p$. The dividing surface predicted by the corresponding optimized coordinate is shown in blue.}
	\label{fig:new_mirrored_fe}
\end{figure}

Before discussing further analyses of the solvent coordinates, we present the two-dimensional free energy landscape for two solute coordinates $r_J$ and the Walden inversion coordinate $r_p$.  The free energy surface $F(r_J, r_p)$, shown in Fig.\,\ref{fig:new_mirrored_fe}, was obtained from umbrella sampling data and Eq.\,\eqref{eq:cond_prob}.  Similar to the one-dimensional projection, a clear saddle separating the two stable basins left and right can be seen, with a barrier height of 14\,kcal/mol, which is very similar to the one-dimensional case. The free energy surface shows that $r_p$ is important near the transition state, but $r_J$ is a better coordinate at early and late stages of the reaction pathway.

Even though the free energy surface $F(r_J, r_p)$ shows a clear saddle with unstable direction pointing largely along $r_p$, the transmission coefficient for coordinate $r_p$ is not better than that for $r_J$. This may be related to the sharp bend in the reaction pathway: a transition path moving along $r_p$ only falls a few $k_{\text{B}} T$ from the barrier top before it has to make a sharp turn moving along the slower $r_J$ coordinate toward the reactant or product state. This may cause $r_p$ to be have dynamics like the non-adiabatic frozen solvent regime, except that the slow coordinate in this case is the solute coordinate $r_J$. The transmission coefficient of the combined ($r_J$, $r_p$) coordinate does not show a statistically significant difference to that of either single coordinate either.

The free energy barrier along $r_p$ was not computed. From the joint free energy landscape for ($r_J$, $r_p$) we can anticipate that the free energy along $r_p$ will give only a small barrier because the edge of the reactant and product basins in Fig.\,\ref{fig:new_mirrored_fe} overlap with the transition state location at $r_p = 0$. In this sense the joint free energy landscape confirms the inertial likelihood prediction that $r_J$ is a better individual reaction coordinate than $r_p$.    

\begin{figure}
	\centering
	\includegraphics[width=1.0\columnwidth]{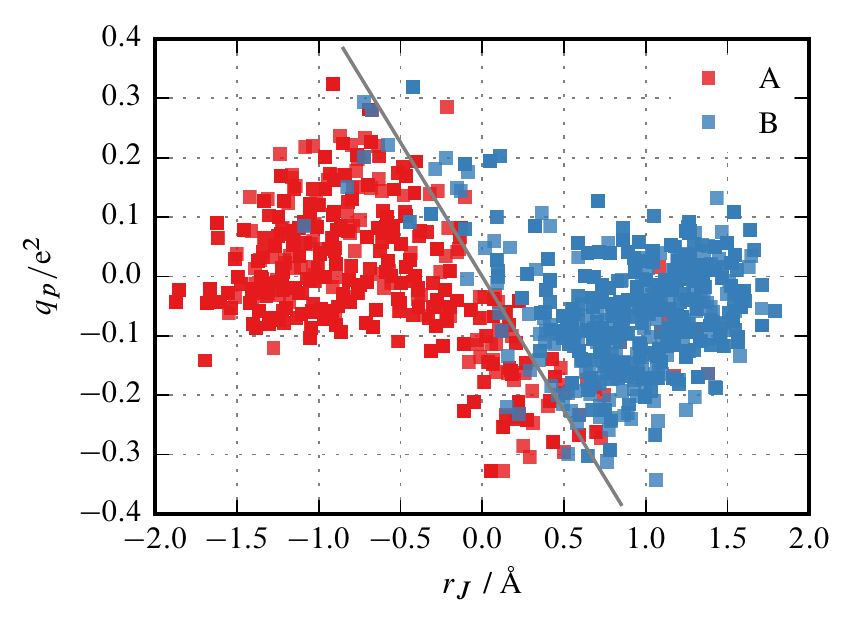}
	\caption{Outcome of forward trajectory parts used in likelihood maximization as a function of $r_J$ and the solvent coordinate $q_p$. The dividing surface predicted is shown in grey. The boundaries used to define the states $A$ and $B$ coincide with the left and right limits of the plot.}
	\label{fig:fwd_outcome}
\end{figure}

Now we examine results for the new solvent coordinate.  In Fig.\,\ref{fig:fwd_outcome} we have plotted the outcome of the forward trajectory parts used in the likelihood maximization as a function of solute and solvent coordinate. When a trajectory started from a specific point ends on the right-hand ($B$) side of the barrier, we plot the corresponding data point in blue. Conversely, for trajectories ending on the left-hand ($A$) side of the barrier, the data point is colored in red. In the same figure, we show the dividing surface predicted by the likelihood maximization procedure for the linear combination of the two variables $r_J$ and $q_p$. Note that most red points are to the left of the dividing surface, while most blue points are to the right of it. Clearly, the solute coordinate is still the most important factor in determining where a trajectory ends: most trajectories that start on the $A$-side of the barrier also end there, and the same is true for the $B$-side of the barrier. However, exceptions occur for trajectories with more extreme values of the solvent coordinate. In these cases, the solvent polarization has enough influence to push the reaction over the barrier. For reference, we have included the likelihood score for the solvent coordinate $q_p$ alone in Table~\ref{tab:likelihoods} as well. 

\begin{figure}
	\centering
	\includegraphics[width=1.0\columnwidth]{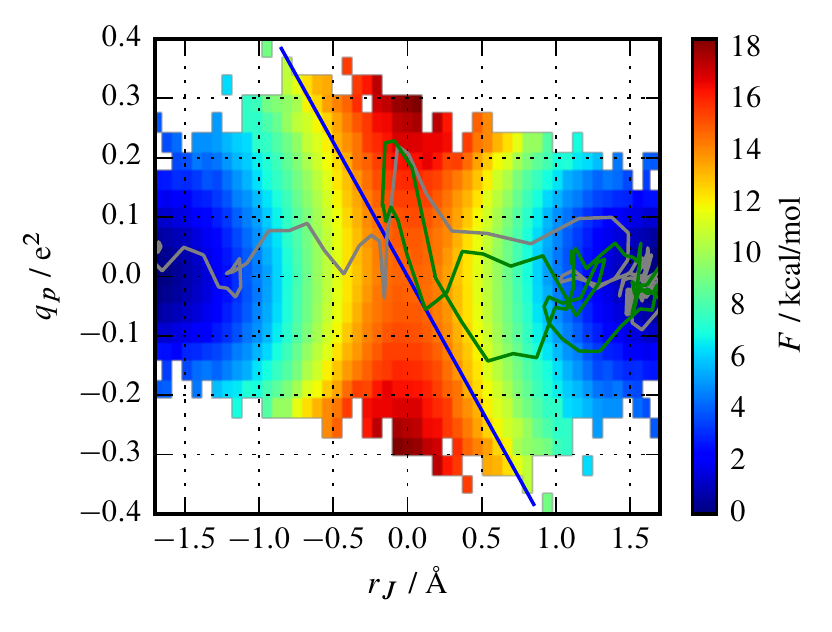}
	\caption{Free energy as a function of $r_J$ and $q_p$. The dividing surface predicted by the corresponding optimized coordinate is shown in blue. Also shown is a randomly selected reactive trajectory (grey) and a nonreactive trajectory (green).}
	\label{fig:new_qp_fe}
\end{figure}

The new solute coordinate $q_p$ is constructed with the general structure suggested by Eq.\,2.1 in Gertner et al.\cite{Gertner1987} Specifically, $q_p$ is defined by deviation from an equilibrium average over solvent degrees of freedom with the solute coordinate fixed. By definition $q_p$ quantifies instantaneous fluctuations from the average at a fixed solute coordinate value, so the solvent coordinate alone cannot perform in a satisfactory manner. Instead, $q_p$ should rather be understood as an augmentation to the solute coordinate. This is also evident from the free energy landscape as a function of these two coordinates (Fig.\,\ref{fig:new_qp_fe}): while there is some structure related to $q_p$ in the free energy, the overall landscape is still dominated by the solute coordinate, especially far from the transition state. We also plot the same dividing surface (blue line) predicted by the likelihood maximization as shown in Fig.\,\ref{fig:fwd_outcome}, as well as two randomly selected trajectories from the TPS simulation. Note how the non-reactive trajectory, plotted in green, never crosses this dividing surface, despite going from positive to negative $r_J$ values and back again.

The free energy along $q_p$ alone was not computed. First, the $q_p$ coordinate alone has a very low likelihood score. Second, the joint free energy landscape for ($r_J$, $q_p$) suggests that that $F(q_p)$ will not even exhibit a barrier, because the reactant, transition state, and product in Fig.\,\ref{fig:fwd_outcome} all have $q_p = 0$. In this sense, $r_J$ is a better individual reaction coordinate than $q_p$. 

\begin{figure}
	\centering
	\includegraphics[width=1.0\columnwidth]{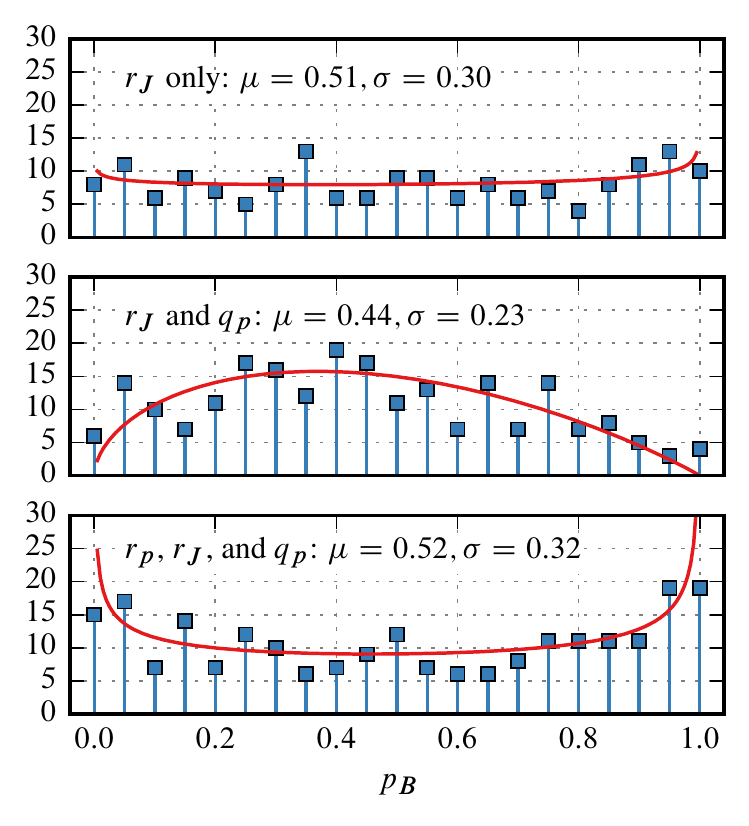}
	\caption{Committor histogram test for the original coordinate $r_J$ and two combined coordinates with a high likelihood score. The actual observed histogram is shown in blue, while in red we have plotted the corresponding $\beta$ distribution with mean and standard variance extracted from the observed discrete distribution.\cite{Peters2006a}}
	\label{fig:new_pbhist_three}
\end{figure}

Results of committor tests are shown in Fig.\,\ref{fig:new_pbhist_three}. Means and standard deviations $(\mu, \sigma)$ of the true committor distribution estimated using Eqs.\,\eqref{eq:mu} and \eqref{eq:sigma} are shown in the figure as well. The solid red lines are $\beta$ distributions with the corresponding means and standard deviations, representing the best estimate for the true continuous committor distribution. 

The committor histogram for the base coordinate $r_J$ is shown in the top panel of Fig.\,\ref{fig:new_pbhist_three}. It shows the approximate committor values of 170 configurations within $-0.01$\,Å$\,\leq r_J \leq 0.01\,$Å. The histogram is rather flat, which is definitely not ideal, but at the same time it is already much better than a bimodal histogram sharply peaked around $p_B = 0$ and $p_B = 1$, which one would expect for a particularly bad reaction coordinate.  While the rather flat histogram observed for $r_J$ is already a decent baseline, we can see a clear improvement for the combination $r_J$--$q_p$. The addition of plane-inversion solute coordinate $r_p$ (bottom panel in Fig.\,\ref{fig:new_pbhist_three}) does not improve the committor histogram, most likely due to the inclusion of many states far from the actual barrier. This has to be contrasted with the high log likelihood score for that very combination, which at first glance seems to be a contradiction. However, the configurations used for the likelihood maximization procedure are shooting points from a TPS simulation, that are on average close to the transition barrier as defined by $p_B = 1/2$. In other words, $r_p$ is useful in predicting the outcome of a trajectory only close to the barrier. Conversely, far from the barrier, a configuration will have a committor value close to 0 or 1 (and a value of $r_J$ that is very different from 0), but (as indicated by the free energy landscape, Fig.\,\ref{fig:new_mirrored_fe}) purely from thermal fluctuations might still have a value of $r_p \approx 0$.

Many of the trial reaction coordinates tested in this work yielded mixed results. For example, the combined ($r_J$, $r_p$, and $q_p$) coordinate resulted in a significantly improved log likelihood score, but did not perform well in a committor test. The combined ($r_J$, $q_p$) coordinate gave a significantly improved likelihood score \textit{and} a better description of the transition state ensemble according to the committor test. However, the combined ($r_J$, $q_p$) coordinate did not improve the transmission coefficient. In fact, we find that $r_J$, $r_p$, and combined ($r_J$, $q_p$) coordinates all have similar transmission coefficients. The failure to improve transmission coefficients with improved coordinates in this work may be ascribed to any of several factors. First, it may be impossible to find a perfect dividing surface for this system as found in work on NaCl dissociation.  Second, the Jorgensen coordinate already gives a transmission coefficient of 0.39, leaving only marginal room for improvement. Third, we conjecture that committor tests are more sensitive indicators of reaction coordinate error than rate calculations and transmission coefficients, but more work is needed to quantify and test the last supposition.

\section{Conclusions}
\label{sec:conclusions_finkelstein}

Using a prototypical S$_{\text{N}}$2 example reaction, we have introduced a new solvent coordinate for chemical reactions in solution. At each point along a solute coordinate, the new solvent coordinate quantifies deviation of the instantaneous charges on solute atoms from their averages at the same point along the solute coordinate. An inner product is constructed from the vectors of instantaneous charge deviations and the net changes in charge between reactants and products. In this way, the new coordinate quantifies whether instantaneous solvent induced charge polarization is pulling along or against the charge transfer direction. The new coordinate can be constructed with Mulliken or Bader charges and with any solute, making it applicable in QM/MM studies of essentially any reaction in solution.

In combination with a suitable solute coordinate, likelihood and committor analyses show that the new solvent coordinate provides an improved description of the reaction dynamics. However, the new solvent coordinate did not improve the transmission coefficient, suggesting that (for purposes of computing a rate) the solute coordinate used in earlier studies was already sufficient.

Nevertheless, our analysis has identified the new solvent coordinate as the most significant contribution beyond $r_J$, more important than coordination numbers and earlier polarization coordinates based on prescribed charges. Even though it did not improve the transmission coefficient in this system, the new coordinate is worth reporting because: (1) such induced polarization coordinates have not been discussed before, (2) on the basis of likelihood scores and committor distributions, it outperforms solvent coordinates that have been used, and (3) it can be applied without modification to any reaction in solution. Therefore the new solvent coordinate may be useful for understanding solvent effects in other systems or for additional studies of the \ce{Cl- + CH3Cl} reaction with better QM/MM models.

\section{Supplementary Material}

In the Supplementary Material, we show how the chosen exchange-correlation functional strongly influences the computed barrier height in gas-phase calculations. Average hydrogen charges along $r_J$ are shown as well. We also give a brief explanation of the EPF procedure used to calculate transmission coefficients and show a few example trajectories from such a calculation. Furthermore, we list log likelihood scores for additional variable combinations not included in Tab.\,\ref{tab:likelihoods}.

\begin{acknowledgments}
This work was supported the U.S. Department of Energy’s (DOE) Office of Basic Energy Sciences, Division of Chemical Sciences, Geosciences and Biosciences. Pacific Northwest National Laboratory (PNNL) is operated for the Department of Energy by Battelle. We acknowlege computer resources through PNNL’s institutional computing (PIC), all the simulations were run on PNNL's \textit{Constance} cluster. BP and CL acknowledge financial support from the National Science Foundation Award No. 1465289 in the Division of Theoretical Chemistry. CJM and GKS were supported the U.S. Department of Energy’s (DOE) Office of Basic Energy Sciences, Division of Chemical Sciences, Geosciences and Biosciences. MDB was supported by the BES Division of Materials Science and Engineering, Synthesis and Processing Sciences Program.
\end{acknowledgments}

\section*{Availability of data}

The data that support the findings of this study are available from the corresponding author upon reasonable request.

\appendix

\section{Calculation of the plane-inversion coordinate}
\label{ap:plane-rc}

In order to define the plane-inversion coordinate $r_p$, one must first find the normal vector of the plane defined by the three hydrogen atoms. This is done with a cross product of two vectors in the plane, e.\,g.
\begin{align}
\vec{d}_1 &= \vec{r}_{H_2} - \vec{r}_{H_1},\\
\vec{d}_2 &= \vec{r}_{H_3} - \vec{r}_{H_1},\\
\vec{n}   &= \vec{d}_1 \times \vec{d}_2.
\end{align}
Now, compute another vector pointing from a point in the plane to the central carbon atom, and then calculate the projection along the plane's normal:
\begin{align}
\vec{d'} &= \vec{r}_{C} -  \vec{r}_{H_1},\\
\vec{d} &= \vec{n} \left(\vec{n} \cdot \vec{d'}\right).
\end{align}
Note that the projected vector does not depend on the orientation of $\vec{n}$, so the choice and order of reference points in the plane does not matter. The reaction coordinate is finally calculated as
\begin{equation}
r_p = \frac{\vec{d} \cdot \vec{r}}{|\vec{r}|},
\end{equation}
where $\vec{r} = \vec{r}_{Cl_2} - \vec{r}_{Cl_1}$ is defining the “positive” and “negative” direction.

\section{Acceleration along Jorgensen's coordinate}

The acceleration along $r_J$ is given as
\begin{equation}
\ddot{r}_J = \frac{\ddot{\vec{r}}_{10} \cdot \vec{r}_{10}}{r_{10}} - \frac{\ddot{\vec{r}}_{20} \cdot \vec{r}_{20}}{r_{20}} + \sum_{i=0}^2 \frac{d}{dt} \nabla_i r_J \cdot \dot{\vec{r}}_i.
\end{equation}
Here, $\nabla_i$ is the gradient with respect to the Cartesian coordinates of particles $C$, $Cl_1$, and $Cl_2$, respectively. The relative position vectors are $\vec{r}_{10} = \vec{r}_{Cl_1} - \vec{r}_{C}$ and $\vec{r}_{20} = \vec{r}_{Cl_2} - \vec{r}_{C}$.

\end{document}